\documentclass[11pt]{article}
\usepackage{latexsym}  
\usepackage{amssymb}
\usepackage{graphicx}
\usepackage{amsmath}

\newcommand{\be}{\begin{equation}}
\newcommand{\ee}{\end{equation}}
\newcommand{\bea}{\begin{eqnarray}}
\newcommand{\eea}{\end{eqnarray}}
\newcommand{\bref}[1]{(\ref{#1})}
\begin{document}
\begin{titlepage}
\begin{flushright}
\today
\end{flushright}
\vspace{4\baselineskip}
\begin{center}
{\Large\bf  Hierarchical masses and universal mixing in quark-lepton mass matrices.}
\end{center}
\vspace{1cm}
\begin{center}
{\large Takeshi Fukuyama$^{a,}$
\footnote{E-mail:fukuyama@se.ritsumei.ac.jp}}
and
{\large Hiroyuki Nishiura$^{b,}$
\footnote{E-mail:nishiura@is.oit.ac.jp}}
\end{center}
\vspace{0.2cm}
\begin{center}
${}^{a}$ {\small \it Research Center for Nuclear Physics (RCNP),
Osaka University, Ibaraki, Osaka, 567-0047, Japan}\\[.2cm]

${}^{b} $ {\small \it Faculty of Information Science and Technology, 
Osaka Institute of Technology,\\ Hirakata, Osaka 573-0196, Japan}

\vskip 10mm
\end{center}
\vskip 10mm
\begin{abstract}
The universality hypothesis for quark and lepton mixing matrices (CKM and MNS) is further developed.
This hypothesis explains why the CKM is almost diagonal whereas the MNS is almost maximally mixed.
If this hypothesis is true, the Dirac CP violating phase of the MNS mixing matrix is bounded around $\pi$ or $0$.
Quark-lepton mass matrices which realizes this hypothesis are constructed, showing simple power law relations among
mass matrices for up-type, down-type quarks and neutrinos.

\end{abstract}
\end{titlepage}
The charged quark and lepton mass matrices have hierarchical structures in different orders. The mixing matrices of quarks and leptons are also quite different. That is, the Cabbibo-Kobayashi-Maskawa (CKM) quark mixing matrix is almost diagonal, whereas the Maki-Nakagawa-Sakata (MNS) lepton mixing matrix is almost maximally mixed.
Concerning with the latter property, we have proposed the universal mixing hypothesis to explain this apparently different structure in mixing matrices.
When we discuss on the relations between quarks and leptons we consider that quarks-leptons belong to a multiplet implicitly like GUT. That is, we cannot take rebasing, independently on quark and lepton sectors. Taking this into consideration,  we consider the base where charged lepton mass matrix is diagonal.
In this base the universal mixing hypothesis asserts \cite{FN1} that (left-handed) unitary matrices for  up-type and down-type quark mass matrices have the same forms up to phase, 
\be
U_u=P^\dagger U_{MNS},~~U_d= U_{MNS}.
\label{proposition1}
\ee
So the CKM quark mixing matrix  ($\equiv V_{CKM}$) is represented by
\be
V_{CKM}= U_{MNS}^\dagger P U_{MNS},
\label{CKM}
\ee
where
\be
P\equiv \left(
\begin{array}{ccc}
e^{i\phi_1}&0&0\\
0&e^{i\phi_2}&0\\
0&0&1\\
\end{array}
\right).
\label{P}
\ee
When we proposed this hypothesis  the lepton mixing angle $\theta_{13}$ in $U_{MNS}$ was undetermined experimentally at that time.
So using free parameters $\phi_1.~\phi_2,~\theta_{13}$, we showed that 

1. Eq. \bref{CKM} with the observed $U_{MNS}$ reproduces the observed $V_{CKM}$, and

2. The $\theta_{13}$  was predicted to be $0.036<s_{13}<0.048$, 
where $s_{13}\equiv \mbox{sin}\theta_{13}$.\\
Unfortunately, the predicted $s_{13}$ is too small for the afterward observed value $\sqrt{0.024}=0.155$ \cite{Daya-Bay}. 

So we modified the hypothesis \bref{proposition1} to \cite{FN2}
\be
U_u=P^\dagger U_{MNS}(\delta^\prime),~~U_d= U_{MNS}(\delta),
\label{proposition2}
\ee
and
\be
V_{CKM}= U_{MNS}(\delta^\prime)^\dagger P U_{MNS}(\delta).
\label{CKM_improved}
\ee
That is, the Dirac CP violating phases in $U_{MNS}$ are assumed to be different in $U_u$ and $U_d$. The diagonal phase matrix $P$ is given by (\ref{P}). 

Let us see how this hypothesis reproduces the observed CKM matrix.
The lepton mixing matrix has the standard form
\begin{equation}
U_{MNS}(\delta)=
\left(
\begin{array}{ccc}
c_{13}c_{12},&c_{13}s_{12},& s_{13}e^{-i\delta}\\
-c_{23}s_{12}-s_{23}c_{12}s_{13}e^{i\delta},&c_{23}c_{12}-s_{23}s_{12}s_{13}e^{i\delta},&s_{23}c_{13}\\
s_{23}s_{12}-c_{23}c_{12}s_{13}e^{i\delta},&-s_{23}c_{12}-c_{23}s_{12}s_{13}e^{i\delta},&c_{23}c_{13}
\end{array}
\right) .
\label{mixing}
\end{equation}
We adopted the following global best fit values by Fogli et al. \cite{Fogli} 
for the mixing angles in (\ref{mixing})
\bea
s_{12}=\sqrt{0.31},~s_{23}=\sqrt{0.39},~s_{13}=\sqrt{0.024}.
\eea
Thus free parameters in new \bref{CKM_improved} are $\phi_1$, $\phi_2$, $\delta$, and $\delta^\prime$. Note that $\delta$ and $\delta^\prime$ are restricted to be near to each other since the difference is considered to be due to renormalization group equation(RGE) effect. 
In contrast with mass matrices, the mixing matrix is affected little by RGE \cite{Fuku} \cite{R-S}. For instance, only A parameter of Wolfenstein parameter changes by roughly $10\%$ from $M_Z$ to $M_{GUT}$.
Substituting these values into \bref{CKM_improved} and using the following four observed values\cite{PDG} of the CKM matrix elements,
\begin{eqnarray}
|(V_{CKM})_{us}|&=&0.2252 \pm 0.0009,\label{CKMus}\\
|(V_{CKM})_{cb}|&=&0.0409 \pm 0.0011,\label{CKMcb}\\
|(V_{CKM})_{ub}|&=&0.00415 \pm 0.00049,\label{CKMub}\\
|(V_{CKM})_{td}|&=&0.0084 \pm  0.0006,\label{CKMtd}
\end{eqnarray}
two allowed regions in the $\delta$-$\delta^\prime$ parameter plane which are consistent with the observed CKM are obtained only for the parameter sets \\

(a) $\phi_1=26.9^\circ$ and $\phi_2=-3.8^\circ$   \hspace{0.5cm}
and \hspace{0.5cm} (b) $\phi_1=-29.9^\circ$ and $\phi_2=-3.8^\circ$. \\ \\
\noindent
The allowed region in the $\delta$-$\delta^\prime$ plane for the case (a) is presented in Fig. 1, in which $\delta$ and $\delta^\prime$ are restricted around $198^\circ$ and $180^\circ$, respectively.
That for the case (b) is presented in Fig. 2, in which $\delta$ and $\delta^\prime$ are restricted around $-18^\circ$ and $0^\circ$, respectively. Namely we find  that the observed CKM elements are consistent with  both  the case (a) and the case (b) in our universal mixing model. In the previous paper \cite{FN2}, we discussed only the case (a) by adopting the result of the global best fit value $\delta=1.1\pi$ by Fogli et al. 
In this letter we have scanned over the whole regions more exhaustively
and found that another region of (b)
is also consistent with the observed CKM in the universal mixing scenario.\\

\begin{figure}[t]
\centering
\includegraphics[width=0.5\textwidth]{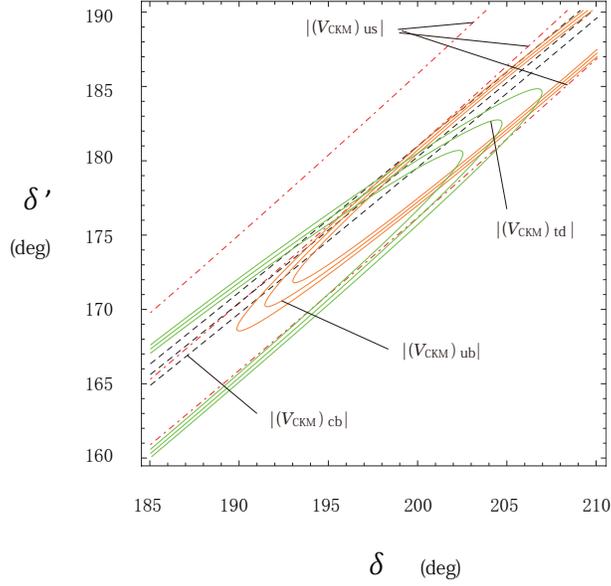}
\caption{Contour plots in the $\delta-\delta^\prime$ plane 
of the observed CKM elements for the case of (a) $\phi_1= 26.9^\circ$ and $\phi_2 = -3.8^\circ$.  
Contour curves for center, lower, and upper values of $|(V_{CKM})_{us}|$, $|(V_{CKM})_{cb}|$, $|(V_{CKM})_{ub}|$, and $|(V_{CKM})_{td}|$ given by \bref{CKMus} - \bref{CKMtd} as functions of $\delta$ and $\delta^\prime$ are drawn by dot-dashed, dashed, light solid(orange), and dark solid(green) curves, respectively. The overlapping region of them is consistent parameter region with the observed $V_{CKM}$. }
\label{fig1}
\end{figure}
\begin{figure}[t]
\centering
\includegraphics[width=0.5\textwidth]{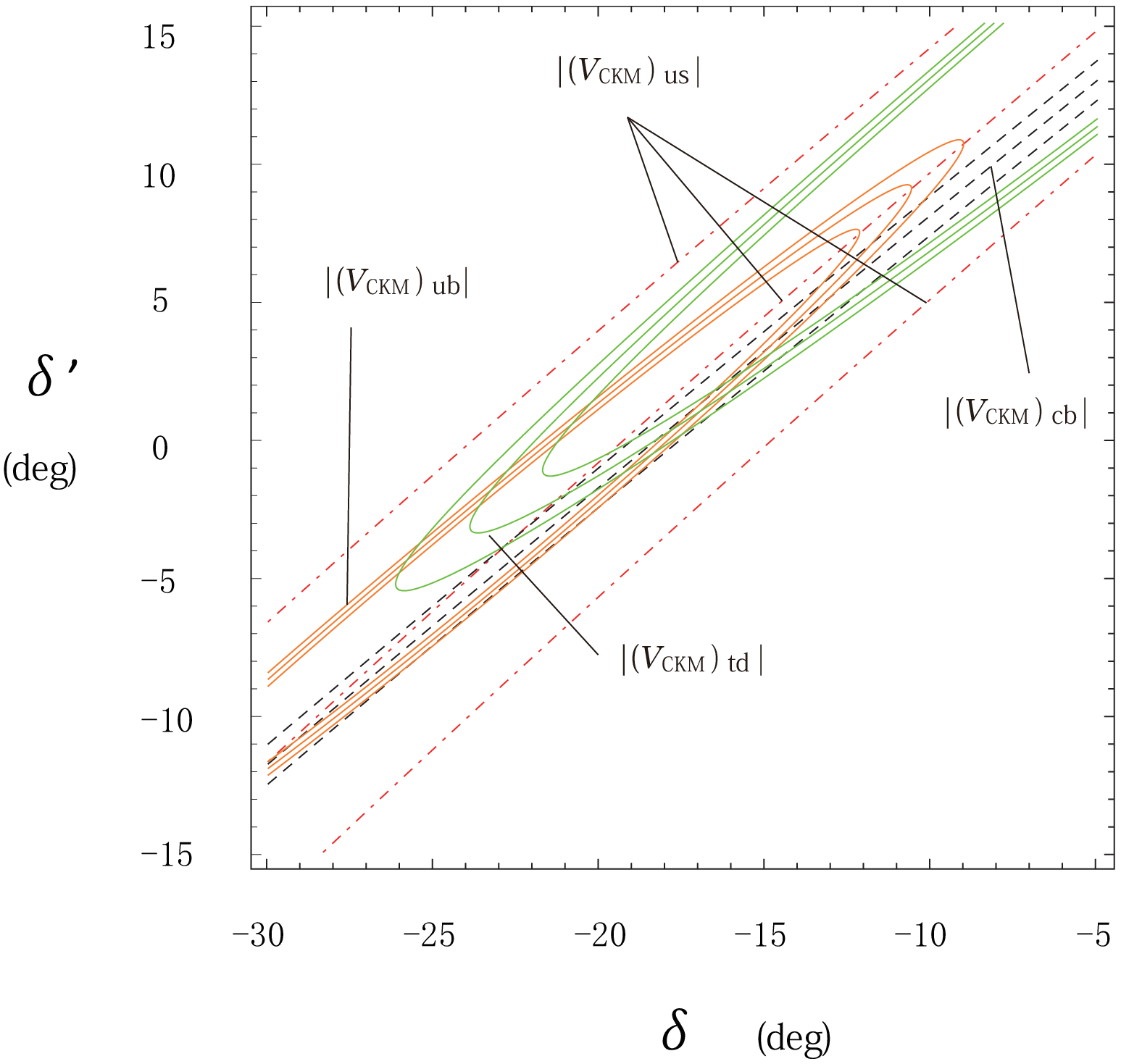}
\caption{Contour plots in the $\delta-\delta^\prime$ plane 
of the observed CKM elements for the case of (b) $\phi_1= -29.9^\circ$ and $\phi_2 = -3.8^\circ$.  
}
\label{fig2}
\end{figure}

So far we have discussed only the universal property of mixing matrices among quarks and
leptons. Now we will discuss how these universalities of mixing matrices are compatible with hierarchical property of mass matrices. 
In our universal mixing model, the mass matrices for up quarks, and down quarks, and Majorana light neutrinos $M_u$,$M_d$, and $M_\nu$ are presented in the diagonal base of $M_e$ as 
\begin{eqnarray}
M_u&=&P^\dagger U_{MNS}(\delta^\prime) 
\left(
\begin{array}{ccc}
m_u & 0& 0\\
0& m_c & 0\\
0 & 0 & m_t\\
\end{array}
\right) 
U_{MNS}(\delta^\prime)^\dagger P \\
M_d&=& U_{MNS}(\delta) 
\left(
\begin{array}{ccc}
m_d & 0& 0\\
0& m_s & 0\\
0 & 0 & m_b\\
\end{array}
\right) 
U_{MNS}(\delta)^\dagger \\
M_\nu &=& U_{MNS}(\delta) 
\left(
\begin{array}{ccc}
m_1 & 0& 0\\
0& m_2 & 0\\
0 & 0 & m_3\\
\end{array}
\right) 
U_{MNS}(\delta)^T ,
\end{eqnarray}
where $m_u$, $m_c$, and $m_t$ are masses of up quarks, $m_d$, $m_s$, and $m_b$ are those of down quarks, 
and $m_1$, $m_2$, and $m_3$ are those of neutrinos. 
Here we have assumed that $M_u$ and $M_d$ are Hermitian and $M_\nu$ symmetric matrices.
It is rather trivial to modify for the other cases. 
Now let us consider $3\times 3$ mass matrix $\Phi$ defined by
\be
\Phi=U_{MNS}(\delta^\prime)\left(
\begin{array}{ccc}
m_u^{1/3}& 0& 0\\
0& m_c^{1/3} & 0\\
0 & 0 & m_t^{1/3}\\
\end{array}
\right) U_{MNS}(\delta^\prime)^\dagger.
\ee
Then, it is interesting that $M_u$, $M_d$, and $M_\nu$ can be described by using common matrix $\Phi$,  as
\begin{eqnarray}
\label{phi3}
&&M_u=P^\dagger \Phi\Phi\Phi P,\\
\label{phi2}
&&M_d=\alpha U_X\Phi\Phi U_X^\dagger +m_{0d}I,\\
&&M_\nu =\beta U_X \Phi U_X^\dagger U_Y + m_{0\nu}U_Y.
\label{neutrino}
\end{eqnarray}
Here $\alpha$, $\beta$, $m_{0d}$, $m_{0\nu}$ are real constants defined by $\alpha =m_b/(m_t)^{2/3}$, $\beta  =m_3/(m_t)^{1/3}$, 
$m_{0d} =m_d-m_b(m_u/m_t)^{2/3}$, and $m_{0\nu} =m_1-m_3(m_u/m_t)^{1/3}$. The mixing matrix $U_X$ appeared in the first term of (\ref{phi2}) and $U_Y$ in (\ref{neutrino}) are respectively given by 
\begin{eqnarray}
U_X&=&U_{MNS}(\delta)U_{MNS}(\delta^\prime)^\dagger \nonumber \\
&=&\left(
\begin{array}{ccc}
1+ \frac{-3+2 \sqrt{2}}{6}(1-e^{i (\delta -\delta^\prime)}) & \frac{e^{-i \delta }-e^{-i \delta^\prime}}{6 \sqrt{2}} & \frac{e^{-i \delta }-e^{-i \delta^\prime}}{6 \sqrt{2}} \\
 \frac{-e^{i \delta }+e^{i \delta^\prime}}{6 \sqrt{2}} & 1+\frac{-3+2 \sqrt{2}}{12}(1-e^{i (\delta -\delta^\prime)}) & \frac{-3+2 \sqrt{2}}{12}(1-e^{i (\delta -\delta^\prime)}) \\
 \frac{-e^{i \delta }+e^{i \delta^\prime}}{6 \sqrt{2}} & \frac{-3+2 \sqrt{2}}{12}(1-e^{i (\delta -\delta^\prime)}) & 1+\frac{-3+2 \sqrt{2}}{12}(1-e^{i (\delta -\delta^\prime)}) \\
\end{array}
\right), \nonumber \\
\mbox{ }\\
U_Y&=&U_{MNS}(\delta) U_{MNS}(\delta)^T\nonumber \\
&=&\left(
\begin{array}{ccc}
1+ \frac{-3+2 \sqrt{2}}{6}(1-e^{i 2\delta }) & \frac{e^{-i \delta }-e^{i \delta}}{6 \sqrt{2}} & \frac{e^{-i \delta }-e^{i \delta}}{6 \sqrt{2}} \\
 \frac{-e^{i \delta }+e^{-i \delta}}{6 \sqrt{2}} & 1+\frac{-3+2 \sqrt{2}}{12}(1-e^{i 2\delta)}) & \frac{-3+2 \sqrt{2}}{12}(1-e^{i 2\delta}) \\
 \frac{-e^{i \delta }+e^{-i \delta}}{6 \sqrt{2}} & \frac{-3+2 \sqrt{2}}{12}(1-e^{i 2\delta}) & 1+\frac{-3+2 \sqrt{2}}{12}(1-e^{i 2\delta}), \\
\end{array}
\right),\nonumber \\
\end{eqnarray}
where we have used a representation of the $U_{MNS}$ proposed by \cite{Xing}.
Note that the $U_X$ and $U_Y$ are 2-3 symmetric mixing matrices, and also that $U_X$ becomes unit matrix in the limit of $\delta=\delta^\prime$, and $U_Y$ does so in the limit of $\delta=0$ or $\delta=\pi$.
The second terms in (\ref{phi2}) and (\ref{neutrino}) are tiny modification terms and not essential. 
The mass matrices given by \bref{phi3},  \bref{phi2}, and  \bref{neutrino} are considered at $M_Z$. It is expected that the following mass matrix relations are satisfied at GUT scale, at which we can neglect the small difference between $\delta$ and $\delta^\prime$ and we can set $U_X=1$ and $U_Y=1$ for $\delta=0$ or $\delta=\pi$.
\begin{eqnarray}
M_u&=&P^\dagger \Phi\Phi\Phi P,
\label{phi3GUT}\\
M_d&=&\alpha \Phi\Phi  +m_{0d}I,
\label{phi2GUT}\\
M_\nu &=&\beta \Phi  + m_{0\nu}I.
\label{neutrinoGUT}
\end{eqnarray}

Now let us discuss consistency among hierarchal masses of quarks and neutrinos. In the subsequent arguments, we adopt the approximation from \bref{phi3GUT}, \bref{phi2GUT}, \bref{neutrinoGUT} at $M_Z$.
The observed value for quark masses are
\begin{eqnarray}\label{data1}
& & m_u = 0.00233 \; \text{GeV}, \; \;   m_c=0.677 \; \text{GeV}, 
\; \;  m_t=176 \;\text{GeV},\nonumber \\
& & m_d=0.00469 \; \text{GeV},  \; \;  m_b=3.00 \;\text{GeV},  \\ 
& & m_e=0.000487 \; \text{GeV}, \; \; m_\mu=0.103 \; \text{GeV},
\; \;  m_\tau=1.75 \;\text{GeV}.\nonumber 
\end{eqnarray} 
Here the experimental values at $M_Z$ were extrapolated from low energies 
 to $M_Z$ \cite{Fusaoka-Koide} except for $m_s$.
$m_s$ value is still ambiguous and we obtained $m_s=72.9$ MeV from the model consistency \cite{Fuku}. Using these, the mass relation between \bref{phi3GUT} and \bref{phi2GUT},
\be
\left(\frac{m_s}{m_b}\right)^3=\left(\frac{m_c}{m_t}\right)^2
\label{2/3}
\ee
is satisfied  within 3\%. \bref{2/3} is satisfied in broad data, not restricted to \bref{data1}.
As for the first generation, we have
\be
\left(\frac{m_d}{m_s}\right)^3=2.66\times 10^{-4}\gg \left(\frac{m_u}{m_c}\right)^2=1.18\times 10^{-5}
\ee
and $m_{0d} =m_d-m_b(m_u/m_t)^{2/3}$. Using \bref{2/3} we find that $M_d$ can be described in the form of $\Phi\Phi$ as shown in \bref{phi2GUT}. 

Lastly we proceed to check \bref{neutrinoGUT} using observed values of neutrino mass-squared differences  \cite{PDG}
\be
\Delta m^2_{32}=(2.32^{+0.12}_{-0.08}) \times 10^{-3}~ \mbox{eV}^2,~~\Delta m^2_{21}=(7.50\pm 0.20)\times 10^{-5}~\mbox{eV}^2.
\label{data2}
\ee
If we assume normal hierarchical structure in neutrino masses, we have
\be
\frac{m_2}{m_3}=\sqrt{\frac{\Delta m^2_{21}}{\Delta m^2_{32}}}.
\ee
Then, using the center values of \bref{data2}, Eq.\bref{neutrinoGUT} predicts
\be
\frac{m_2}{m_3}=\sqrt{\frac{\Delta m^2_{21}}{\Delta m^2_{32}}}=1.8\times 10^{-1} \simeq \left(\frac{m_c}{m_t}\right)^{1/3}=1.6\times 10^{-1}.
\ee
If we take the adopted approximations into consideration, 
the consistency is not so bad.

Thus the universality hypothesis predicts very small CP phase effect in neutrino oscillation.
Mass matrices which realize this hypothesis show simple power law relations among $M_u$,$M_d$, and $M_\nu$.

\section*{Acknowledgements}
The work of T.F.\ is supported in part by the Grant-in-Aid for Science Research
from the Ministry of Education, Science and Culture of Japan
(No.~21104004).

\end{document}